\documentclass{article} 

 \usepackage[utf8]{inputenc} 
 \usepackage[T1]{fontenc}    
\usepackage[colorlinks=true, allcolors=blue]{hyperref}
 \usepackage{url}            
 \usepackage{booktabs}       
 \usepackage{amsfonts}       
 \usepackage{nicefrac}       
 \usepackage{microtype}      
 \usepackage{xcolor}         
 \usepackage{graphicx}       
 \usepackage{amsmath}        
 \usepackage{amssymb} 
 \usepackage{mathtools} 
 \usepackage{amsthm} 

 \usepackage[preprint]{neurips_2025} 
 \workshoptitle{Published in the Proceedings of the 39th Conference on Neural Information Processing Systems (NeurIPS 2025) Workshop: Scaling Environments for Agents (SEA). Additionally accepted for presentation in the NeurIPS 2025 Workshop: Embodied World Models for Decision Making (EWM) and the NeurIPS 2025 Workshop: Optimization for Machine Learning (OPT).} 

 \title{Communicating Plans, Not Percepts: Scalable Multi-Agent Coordination with Embodied World Models} 

\author{%
  Brennen A. Hill\thanks{Corresponding author.}\\
  Department of Computer Science\\
  University of Wisconsin-Madison\\
  Madison, WI 53706 \\
  \texttt{bahill4@wisc.edu} \\
  \And
  Mant Koh En Wei \\
  Department of Computer Science\\
  National University of Singapore\\
  Singapore 119077 \\
  \texttt{e0958776@u.nus.edu} \\
  \And
  Thangavel Jishnuanandh \\
  Department of Computer Science\\
  National University of Singapore\\
  Singapore 119077 \\
  \texttt{jishnuanandh@u.nus.edu} \\
}

\begin{document} 

 \maketitle 

 \begin{abstract} 
 Robust coordination is critical for effective decision-making in multi-agent systems, especially under partial observability. A central question in Multi-Agent Reinforcement Learning (MARL) is whether to engineer communication protocols or learn them end-to-end. We investigate this dichotomy using embodied world models. We propose and compare two communication strategies for a cooperative task-allocation problem. The first, Learned Direct Communication (LDC), learns a protocol end-to-end. The second, Intention Communication, uses an engineered inductive bias: a compact, learned world model, the Imagined Trajectory Generation Module (ITGM), which uses the agent's own policy to simulate future states. A Message Generation Network (MGN) then compresses this plan into a message. We evaluate these approaches on goal-directed interaction in a grid world, a canonical abstraction for embodied AI problems, while scaling environmental complexity. Our experiments reveal that while emergent communication is viable in simple settings, the engineered, world model-based approach shows superior performance, sample efficiency, and scalability as complexity increases. These findings advocate for integrating structured, predictive models into MARL agents to enable active, goal-driven coordination. 
 \end{abstract} 

 \section{Introduction} 

 Embodied artificial intelligence seeks to create agents that understand, predict, and interact with the physical world to achieve complex goals. The ability of multiple agents to coordinate is paramount in applications from autonomous driving to warehouse robotics \citep{vincentMarlRobotics, anglenMarlImpact}. However, multi-agent coordination is a formidable challenge for reinforcement learning. For any single agent, the environment is non-stationary because other agents' actions alter the state transition dynamics, violating the Markov property central to many single-agent algorithms \citep{kefanMarlstationary}. This non-stationarity makes learning a stable policy or accurate world model exceedingly difficult. 

 Communication can mitigate this challenge. By exchanging information, agents can synchronize knowledge, align intentions, and make their behavior more predictable, restoring a degree of stationarity \citep{yangMarlCommsEffective}. This raises a key question: what is the most effective way to design a communication protocol for decentralized, cooperative agents? 

 One approach is \textit{emergent communication}, where agents learn a protocol from scratch via environmental rewards. This approach is general, requires no hand-engineered features, and may discover novel, efficient encoding schemes. A contrasting approach is \textit{engineered communication}, where protocols have strong inductive biases that structure message content. This approach posits that for complex coordination, a scaffold for reasoning and communication is more sample-efficient and scalable. 

 This paper explores this trade-off using world models. We reframe the debate as a comparison between implicit versus explicit world modeling for decision-making. Can agents implicitly model each other's intentions and future behavior through an unstructured, emergent protocol? Or is it more effective to give agents an explicit world model and engineer communication to share its outputs? 

 To investigate these questions, we focus on a canonical multi-agent coordination problem: cooperative task allocation in a partially observable grid world. Two agents must navigate to two distinct goals, requiring them to resolve ambiguity and avoid conflict. We introduce and compare two approaches: 
 \begin{enumerate} 
     \item \textbf{Learned Direct Communication (LDC):} An end-to-end approach where policy networks produce both an action and a low-bandwidth message. The message's meaning is entirely emergent. 
     \item \textbf{Intention Communication:} An engineered approach where each agent has a lightweight, learned world model, the \textbf{Imagined Trajectory Generation Module (ITGM)}, that performs a short-horizon "mental simulation" of its future trajectory. A Message Generation Network (MGN) then compresses this simulated plan into a message, communicating the agent's intent. 
 \end{enumerate} 

 We investigate how world models can benefit model-based reinforcement learning and long-horizon planning, even at a small scale. We show how a predictive model enables agents to perform active, goal-driven interaction rather than passive reaction. Our results, obtained under computational constraints, underscore the importance of sample efficiency and scalability for real-world deployment. We find that while emergent communication shows some success, the structured, predictive nature of Intention Communication leads to superior performance, robustness, and scalability. This suggests that for complex, embodied coordination, the inductive biases of an explicit world model are essential. 

 \section{Related Work} 

 Our research builds on three areas: emergent communication in MARL, model-based reinforcement learning with world models, and multi-agent planning. 

 \subsection{Emergent Communication in Multi-Agent Reinforcement Learning} 
 Agents learning their own communication protocols has a rich history. Early work like DIAL \citep{foerster2016learning} and CommNet \citep{sukhbaatar2016learning} showed that agents could learn to pass messages to solve cooperative tasks. DIAL showed that gradients could be passed through communication channels, while CommNet introduced a centralized module to pool communications. These approaches treated messages as latent vectors, with the protocol's semantics emerging from optimizing the task reward. RIAL \citep{foerster2016learning} extended this to independent Q-learners. More recent work has explored these emergent languages' properties, such as their compositionality and ability to generalize \citep{mordatch2018emergence}. 

 However, these methods often require high sample complexity to establish a meaningful protocol. The sparse and noisy environmental learning signal makes credit assignment for messages difficult. Furthermore, emergent protocols are often brittle and lack interpretability, making them hard to debug or trust. Our LDC model represents this paradigm, testing the limits of pure emergence in a resource-constrained setting. 

 \subsection{Model-Based Reinforcement Learning and World Models} 
 Model-Based Reinforcement Learning (MBRL) improves sample efficiency by learning a model of the environment's dynamics. This "world model" can then be used for planning, policy improvement, or data augmentation. A prominent research area is learning these world models in a latent space, which is more scalable and robust than predicting raw sensory inputs. 

 The Dreamer family of algorithms \citep{hafner2019dream, hafner2020mastering, hafner2023mastering} are state-of-the-art, learning a recurrent state-space model (RSSM) to predict future latent states and rewards. Policies are then trained entirely within the "dream" of this learned world model. These models have succeeded in single-agent, long-horizon tasks, showing that an accurate world model can replace real environment interaction. 

 Our Intention Communication approach is inspired by this paradigm. The Imagined Trajectory Generation Module (ITGM) is a simplified, computationally inexpensive world model. While less sophisticated than an RSSM, it captures the core principle of learning a predictive model to enable forward-looking decisions. We adapt this to the multi-agent context, positing that a world model's output is useful not only for an agent's own planning but also as an ideal substrate for communication. 

 \subsection{Planning and Intention in Multi-Agent Systems} 
 Effective coordination requires reasoning about others' future actions and intentions. Some MARL approaches explicitly model teammate policies \citep{raileanu2018modeling}, but this can be challenging with many agents or complex policies. 

 An alternative is for agents to communicate their plans directly. This has been explored in classic AI planning and, more recently, deep learning. For example, \citep{liu2021islff} proposed sharing planned trajectories to improve coordination in autonomous driving. By making plans explicit, agents can anticipate and avoid conflicts, simplifying coordination. 

 Our Intention Communication model operationalizes this concept. The ITGM's "imagined trajectory" is a proxy for the agent's short-term plan. By compressing this trajectory into a message, an agent gives its teammate a rich, forward-looking signal far more informative than a single action or goal declaration. This transforms the problem from inferring intent from low-level actions to directly incorporating a teammate's plan into one's own decision-making. This explicit sharing of world model outputs moves beyond passive prediction towards goal-directed interaction. 

 \section{Problem Formulation and Experimental Environment} 

 We formalize our cooperative task as a Decentralized Partially Observable Markov Decision Process (Dec-POMDP) and detail our experimental environment. 

 \subsection{Dec-POMDP Formulation} 
 A Dec-POMDP is a tuple $(\mathcal{I}, \mathcal{S}, \{\mathcal{A}_i\}_{i \in \mathcal{I}}, T, R, \{\Omega_i\}_{i \in \mathcal{I}}, O)$, where: 
 \begin{itemize} 
     \item $\mathcal{I} = \{1, ..., N\}$ is the set of $N$ agents. In our case, $N=2$. 
     \item $\mathcal{S}$ is the set of environment states. A state $s \in \mathcal{S}$ defines the coordinates of both agents and both goals. 
     \item $\mathcal{A}_i$ is the set of actions available to agent $i$. Agents collectively take a joint action $\mathbf{a} = (a_1, ..., a_N) \in \mathcal{A} = \times_{i \in \mathcal{I}} \mathcal{A}_i$. 
     \item $T(s' | s, \mathbf{a})$ is the state transition function, which is deterministic in our setup. 
     \item $R(s, \mathbf{a})$ is the shared reward function for the team. 
     \item $\Omega_i$ is the set of observations for agent $i$. 
     \item $O(o | s, \mathbf{a})$ is the observation function, determining the probability of agent $i$ receiving observation $o_i \in \Omega_i$ after joint action $\mathbf{a}$ leads to state $s'$. 
 \end{itemize} 
 Each agent $i$ learns a policy $\pi_i(a_i | h_i)$ mapping its history $h_i$ to an action $a_i$ to maximize the expected discounted team reward $E[\sum_{t=0}^{T} \gamma^t R(s_t, \mathbf{a}_t)]$. With communication, the policy becomes $\pi_i(a_i | h_i, m_{t-1})$, conditioned on received messages. 

 \subsection{Experimental Environment} 
 We use a simple, deterministic grid world as a controlled testbed. This canonical environment isolates the communication protocol's effects from confounding factors like complex physics or stochasticity. Its simplicity is a feature, enabling us to probe the challenges of coordination and scalability. 
 \paragraph{Setup} The environment is an $X \times X$ grid containing two agents and two goals, each occupying a unique cell. For our experiments, $X$ is varied to test scalability (e.g., 6, 10, 15). We use the PettingZoo library \citep{terry2021pettingzoo} for its flexible MARL API. All experiments were run on Google Colab, a resource-constrained platform that prioritizes sample efficiency and model compactness. 
 \paragraph{Actions} At each discrete timestep, an agent can choose one of five actions: `stay`, `up`, `down`, `left`, or `right`. 
 \paragraph{Observations} We examine two settings. \textbf{Fully Observable:} Each agent receives the global state: its own coordinates, its teammate's coordinates, and both goal coordinates. To break symmetry and prevent static goal assignments, the order of goal coordinates in the observation vector is randomized for one agent. \textbf{Partially Observable:} Each agent observes its own absolute coordinates and those of any object (teammate or goal) within its \texttt{vision\_range}. This locality makes communication essential for global coordination. 

 To provide minimal memory, the policy network input is a stack of the last four observations. 
 \paragraph{Task and Reward Scheme} The cooperative task is for both agents to occupy the two distinct goals. The shared reward encourages this outcome efficiently: 
 \begin{itemize} 
     \item $\mathbf{+1.0}$ for success (both agents on different goals). 
     \item $\mathbf{-0.10}$ penalty for collision (both agents on the same goal). 
     \item $\mathbf{-0.01}$ per-step time penalty to incentivize efficient paths. 
 \end{itemize} 

 \paragraph{Episode Structure}Each episode begins with agents and goals placed randomly on distinct cells. An episode terminates upon success or after 200 timesteps. 

 \begin{figure}[h] 
     \centering 
     \includegraphics[width=0.3\linewidth]{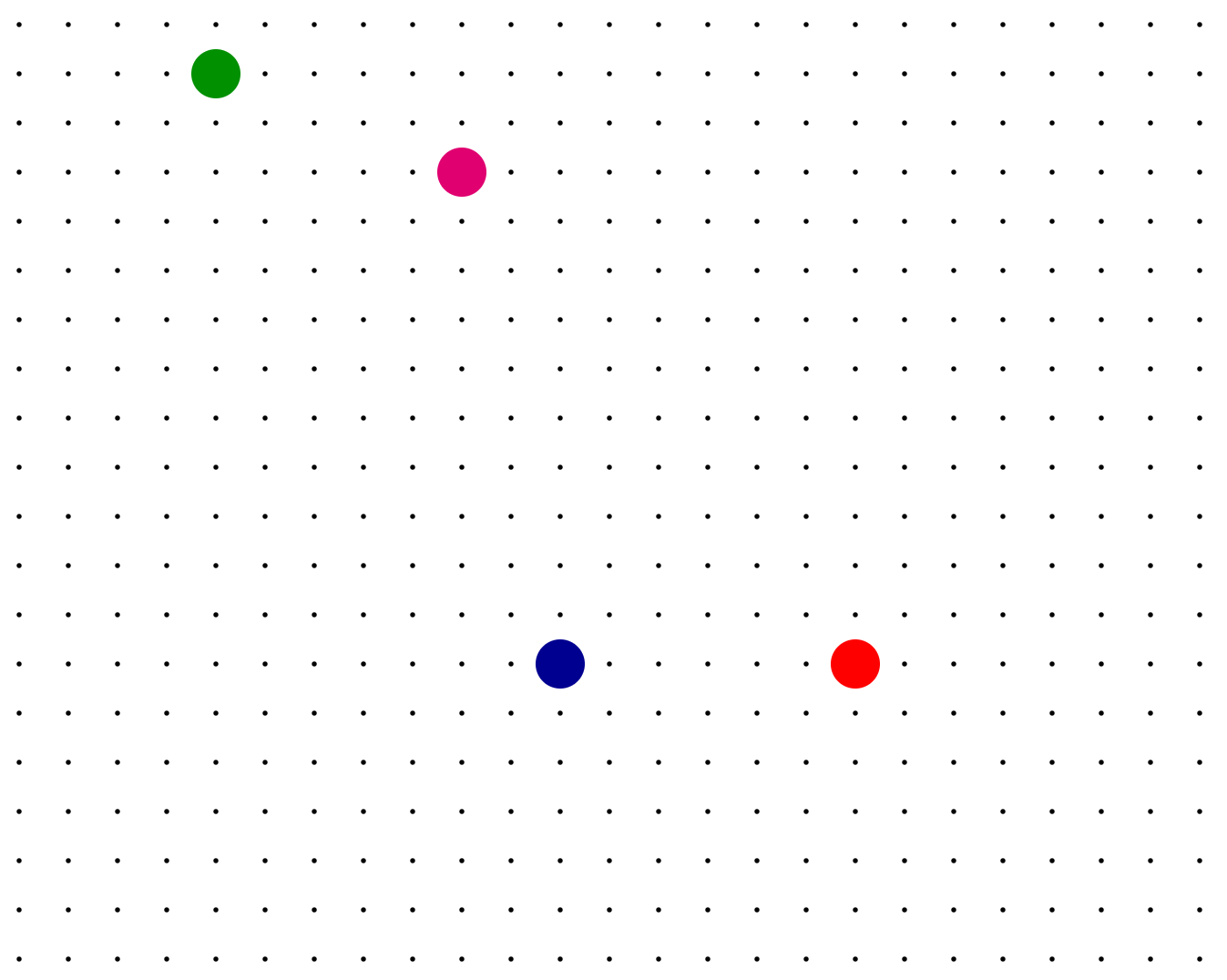} 
     \caption{A visualization of the $6 \times 6$ grid world environment with two agents (blue and red) and two goals (green squares).} 
     \label{fig:board} 
 \end{figure} 

 \subsection{Training Framework} 
 We use an on-policy Advantage Actor-Critic (A2c) algorithm with a learned baseline \citep{sutton2018reinforcement,NIPS1999_6449f44a}. The actor and critic are separate neural networks with shared initial layers. Updates are performed at the end of each episode. We use a linear learning rate decay schedule to balance exploration and exploitation. The learning rate $\text{lr}_e$ for a given episode $e$ is calculated as: 
 \[ 
   \text{lr}_{e} = \text{lr}_0 \left(1 - \frac{e}{E_{\text{total}}}\right) 
 \] 
 where $\text{lr}_0$ is the initial learning rate and $E_{\text{total}}$ is the total number of training episodes. This schedule, clipped at a minimum value of $1 \times 10^{-5}$, allows for larger updates early in training and fine-tuning updates later, preventing catastrophic forgetting. 

 \section{Approach 1: Emergent Communication via Latent Messaging} 

 Our first approach, Learned Direct Communication (LDC), embodies the principle of emergent communication. We create a minimal framework for information exchange, imposing no structure on message content or meaning. The protocol emerges end-to-end, driven only by the shared task reward. 

 \subsection{Architecture and Information Flow} 
\begin{figure}[h]
\centering
\includegraphics[width=1\linewidth]{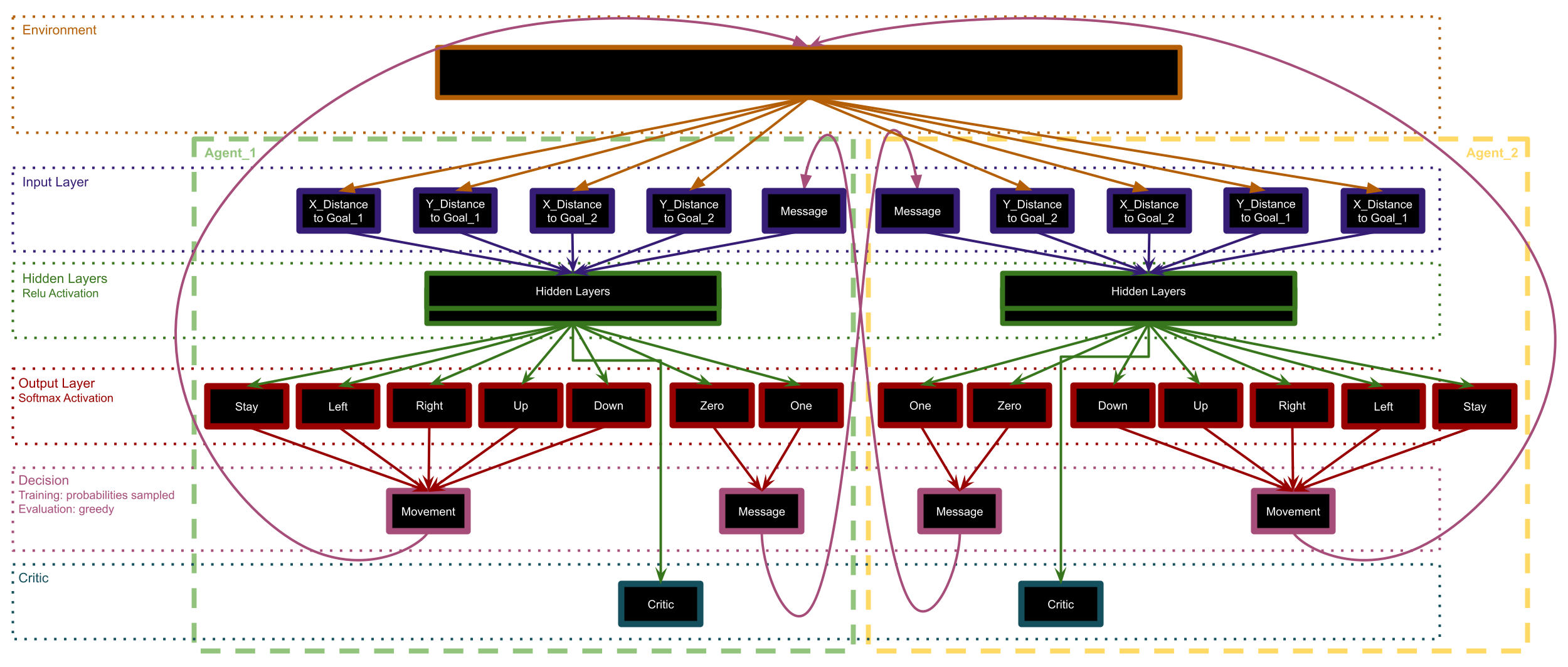}
\caption{The LDC architecture. Each agent's policy network takes its local observation and the incoming message from the previous timestep to produce an action and an outgoing message for the next timestep. The critic network estimates the state value.}
\label{fig:ldc_arch}
\end{figure}

 In LDC, each agent's policy network $\pi_{\theta_i}$ outputs both an action and a message: $\pi_{\theta_i}(a_t^{(i)}, m_t^{(i)} | o_t^{(i)}, m_{t-1}^{(j)})$, where $j \neq i$. We keep the message space minimal to facilitate learning: a single binary value, $m_t^{(i)} \in \{0, 1\}$. The network outputs logits for the action (passed through softmax) and message (passed through sigmoid). During training, both are sampled. The message $m_t^{(i)}$ is broadcast to agent $j$ for use at timestep $t+1$. The entire system is trained end-to-end, with gradients from the shared task reward shaping both the policy and the protocol. The detailed network architecture for Learned Direct Communication is depicted in figure \ref{fig:ldc_arch}. 

 \subsection{Hypothesis and Expected Behavior} 
 The hypothesis is that agents can encode crucial information into the binary message. For instance, '0' might emerge to mean "I am targeting the first goal," and '1' might mean "I am targeting the second." This convention would let agents resolve ambiguity and head to different goals. The approach is flexible; agents can discover any protocol, possibly one more efficient than a human-designed one. The challenge, however, is this unguided discovery process relies on a weak, delayed reward signal to shape the complex joint space of policies and protocols. 

 \subsection{Experimental Analysis in Fully Observable Setting} 
 To validate if LDC could produce meaningful communication, we first tested it in a fully observable $6 \times 6$ grid. With complete goal information, communication's primary role is to break symmetry and coordinate goal selection. 

 \subsubsection{Conditional Probability Study} 
 After training, we analyzed the protocol by examining the conditional probability of an agent's action given a received message. As shown in Tables \ref{tab:conditional1} and \ref{tab:conditional2}, the distributions are clearly distinct. For example, Agent 1 is almost three times more likely to move `Up` when receiving message `0` (13.54\%) compared to message `1` (4.53\%). This strong statistical dependency indicates the messages convey information that systematically influences the recipient's behavior. 

 \begin{table}[h] 
   \caption{Agent 1's Action Probabilities Conditioned on Agent 2's Message} 
   \label{tab:conditional1} 
   \centering 
   \begin{tabular}{lccccc} 
     \toprule 
     & \multicolumn{5}{c}{Action} \\ 
     \cmidrule(lr){2-6} 
     Message & Stay & Left & Right & Up & Down \\ 
     \midrule 
     0 & 31.47\% & 16.98\% & 19.21\% & 13.54\% & 18.80\% \\ 
     1 & 33.95\% & 22.86\% & 14.76\% & 4.53\% & 23.90\% \\ 
     \bottomrule 
   \end{tabular} 
 \end{table} 

 \begin{table}[h] 
   \caption{Agent 2's Action Probabilities Conditioned on Agent 1's Message} 
   \label{tab:conditional2} 
   \centering 
   \begin{tabular}{lccccc} 
     \toprule 
     & \multicolumn{5}{c}{Action} \\ 
     \cmidrule(lr){2-6} 
     Message & Stay & Left & Right & Up & Down \\ 
     \midrule 
     0 & 13.04\% & 21.61\% & 38.08\% & 22.83\% & 4.44\% \\ 
     1 & 13.81\% & 24.55\% & 29.18\% & 24.75\% & 7.71\% \\ 
     \bottomrule 
   \end{tabular} 
 \end{table} 

 \subsubsection{Ablation Study} 
 To establish a causal link, we conducted an ablation study. We took a fully trained LDC policy and, during evaluation, replaced all transmitted messages with a constant value (`0`). As shown in Table \ref{table:ablation-success-rate}, ablating communication caused a discernible drop in success rate and a slight increase in average steps to completion. This confirms the messages are actively used for more effective coordination, not just correlated with behavior. 

 \begin{table}[h] 
   \caption{LDC Performance With and Without Communication (Fully Observable, $6 \times 6$ Grid)} 
   \label{table:ablation-success-rate} 
   \centering 
   \begin{tabular}{lcc} 
     \toprule 
     Condition & Success Rate & Average Steps\\ 
     \midrule 
     With Emergent Messages & 89.4\% & 4.39\\ 
     Messages Ablated (Set to 0) & 88.6\% & 4.43\\ 
     \bottomrule 
   \end{tabular} 
 \end{table} 

 \subsection{Performance under Partial Observability} 
 We then used a more challenging partially observable setting (\texttt{vision\_range} = 3 in a $6 \times 6$ grid). Here, communication is for both coordination and sharing information about the unseen world. As shown in Table \ref{table:partially-observable-success-rate}, the success rate drops for both conditions, as agents must now explore. However, the performance gap widens, with communication providing a more significant relative benefit. This shows that as the need for information sharing increases, so does the emergent protocol's value. 

 \begin{table}[h] 
   \caption{LDC Performance With and Without Communication (Partially Observable, $6 \times 6$ Grid)} 
   \label{table:partially-observable-success-rate} 
   \centering 
   \begin{tabular}{lc} 
     \toprule 
     Condition & Success Rate \\ 
     \midrule 
     With Emergent Messages     & 31.89\% \\ 
     Messages Ablated (Set to 0)  & 30.26\% \\ 
     \bottomrule 
   \end{tabular} 
 \end{table} 

 While LDC demonstrates emergent communication, its performance in simple settings hints at scaling challenges, motivating our second, more structured approach. 

\section{Approach 2: Intention Communication via Latent World Models}
To address the limitations of emergent protocols, we designed Intention Communication, an architecture with strong inductive biases for forward-looking coordination. The core idea is for agents to share explicit information about their future plans, allowing their teammates to anticipate their behavior and avoid conflicts \citep{liu2021islff}. This is operationalized through a two-stage process: an agent first generates an internal mental simulation of its future trajectory, and then compresses this plan into a message for its teammate.

\subsection{Imagined Trajectory Generation Module (ITGM): A Latent World Model}
The ITGM serves as a compact, learned module to generate a plan. It combines a learned latent world model with the agent's own policy to generate a short-term forecast. Its objective is to simulate a future trajectory that would result from following its current policy. This policy-conditioned simulation provides the raw material for forming an intention. The process is as follows.

\paragraph{Initial State Encoding:} At timestep $t$, the module first takes the agent's current local observation $o_t^{(i)}$ (a $(4 \times F)$-dimensional vector with $F=4$) and the received message from its teammate in the previous step, $m_{t-1}^{(j)}$. (This message $m_{t-1}^{(j)}$ is first embedded and flattened into a fixed-length vector). These are concatenated and passed through a linear encoder to produce an initial $D_{hidden}$-dimensional latent state vector, $z_t^{(i)}$: $z_{t}^{(i)} = \text{ReLU}(\mathbf{W}_{enc}[o_t^{(i)} \oplus \text{embed}(m_{t-1}^{(j)})] + \mathbf{b}_{enc})$.

where $\oplus$ denotes vector concatenation. This initial state $z_t^{(i)}$ represents a compressed summary of the agent's current knowledge and serves as the starting point for the plan.

\paragraph{Latent Rollout:} The ITGM generates the plan by unrolling a trajectory for a fixed horizon $H$. This rollout actively uses the agent's own policy head ($\pi_{act}$) and a learned latent transition model ($f_{trans}$). This transition model is a network (e.g., an MLP) that predicts the next latent state given the current latent state and a *planned action*. 

For each step $k$ in the horizon ($k=0, \dots, H-1$), a planned action is sampled from the agent's policy and used by the latent transition model to predict the next state. This iterative process is implemented as follows, where $\text{embed}(\cdot)$ is an action embedding layer and $\pi_{act}$ is the same policy head used to select the agent's real action:
$$\tilde{a}_{t+k} \sim \pi_{act}(\cdot | z_{t+k}^{(i)})$$
$$z_{t+k+1}^{(i)} = \text{ReLU}(\mathbf{W}_{trans}[z_{t+k}^{(i)} \oplus \text{embed}(\tilde{a}_{t+k})] + \mathbf{b}_{trans})$$
This iterative process generates a sequence of predicted future latent states, which forms the plan (the policy-conditioned trajectory): $\tau^{(i)} = \langle z_{t+1}^{(i)}, z_{t+2}^{(i)}, \dots, z_{t+H}^{(i)} \rangle$.

This feed-forward design was chosen for its computational efficiency, which is critical given our resource constraints. Both the encoder, the transition model, and the forward model are feed-forward networks with $D_{hidden}$ hidden units and ReLU activation functions.

\subsection{Message Generation Network (MGN): Summarizing Intentions}
The role of the MGN is to distill the high-dimensional plan (the latent trajectory $\tau^{(i)}$) into a compact, fixed-length message vector $m_t^{(i)}$ that effectively summarizes the agent's intention. A raw sequence of latent states is too unwieldy to use as a direct communication signal. Instead, we use a self-attention mechanism to identify and aggregate the most salient features of the planned trajectory.

We employ a multi-head self-attention mechanism \citep{vaswani2017attention} with $N_{heads}$ head(s) and a $D_{hidden}$-dimensional embedding to compute a weighted summary of the trajectory sequence. The sequence $\tau^{(i)}$ serves as the query, key, and value inputs:$\mathbf{Q} = \tau^{(i)}\mathbf{W}_Q, \quad \mathbf{K} = \tau^{(i)}\mathbf{W}_K, \quad \mathbf{V} = \tau^{(i)}\mathbf{W}_V$. The attention mechanism produces an output sequence where each element is a weighted sum of the values, with weights determined by the query-key similarities. This process allows the network to learn which future steps in its imagined plan are most important for coordination (e.g., the final destination, a potential point of conflict).

This attention output is aggregated via mean pooling to produce a single vector $v_{pooled}$. This vector is then passed through two hidden layers, $h_1 = \text{ReLU}(\mathbf{W}_1 v_{pooled} + b_1)$ and $h_2 = \text{ReLU}(\mathbf{W}_2 h_1 + b_2)$.

To produce the message policy, $h_2$ is fed into $N_{comp}$ separate linear heads. Each head $k \in [1, N_{comp}]$ outputs a vector of $N_{sym}$ logits: $\text{logits}^{(k)} = \mathbf{W}_{head}^{(k)} h_2 + b_{head}^{(k)}$.

This defines $N_{comp}$ independent categorical distributions, $\pi_{msg}^{(k)} = \text{Softmax}(\text{logits}^{(k)})$, one for each component of the message. This structured, multi-part discrete message is far more expressive than the single bit used in LDC.

\subsection{Integration and End-to-End Training}
The ITGM and MGN modules are seamlessly integrated into the agent's decision-making process. The full forward pass for a single agent $i$ at timestep $t$ explicitly separates the action and message generation to respect the constraint:
\begin{enumerate}
\item The agent receives its observation $o_t^{(i)}$ and the previous message from its teammate, $m_{t-1}^{(j)}$.
\item These inputs are encoded to produce the initial latent state $z_t^{(i)}$.
\item (Action Branch) The agent's policy head computes the real action distribution using this initial state: $\pi_{act}(a_t^{(i)} | z_t^{(i)})$. An action $a_t^{(i)}$ is sampled.
\item (Plan/Message Branch) In parallel, the ITGM uses $z_t^{(i)}$ as its starting point to generate the plan $\tau^{(i)} = \langle z_{t+1}^{(i)}, \dots, z_{t+H}^{(i)} \rangle$. This rollout process re-uses the same policy head $\pi_{act}$ to sample hypothetical future actions $\tilde{a}_{t+k}$ at each step of the simulation.
\item The MGN processes the resulting plan $\tau^{(i)}$ to output the $N_{comp} \times N_{sym}$ logits for the $N_{comp}$ message distributions $\pi_{msg}^{(k)}$.
\item A message $m_t^{(i)} = \langle m_t^{(1)}, \dots, m_t^{(N_{comp})} \rangle$ is sampled by taking one symbol from each of the $N_{comp}$ categorical distributions. This composite message vector is broadcast to its teammate for use at step $t+1$.
\item The entire model is trained end-to-end using the A2C loss function. Gradients from the $L_{\text{actor}}$ term (specifically the $A_t \sum_k \log \pi_{msg}^{(k)}(m_t^{(k)})$ component) flow back to update the weights of the MGN, its hidden layers, the ITGM's transition model ($\mathbf{W}_{trans}$), and the initial encoder ($\mathbf{W}_{enc}$). The gradients from both the action ($A_t \log \pi_{act}(a_t)$) and message terms also update the shared policy head $\pi_{act}$.
\end{enumerate}
The entire model is trained end-to-end using the A2C loss function, allowing gradients to flow back through all modules.

\section{Results and Analysis: The Superiority of Engineered World Models} 

 To compare our approaches, we experimented in partially observable environments of increasing scale. These experiments test each strategy's limits on performance, scalability, and sample efficiency, core metrics for embodied agents. Agents had a \texttt{vision\_range} of 2 and a 200-step episode limit. 

 \subsection{Scalability Showdown} 
 We evaluated a non-communicating baseline, LDC, and Intention Communication in $10 \times 10$ and $15 \times 15$ grids. The $15 \times 15$ environment's state space is over 5 times larger, posing a significant challenge. The results, summarized in Table \ref{tab:communication-success-rates}, are stark. 

 \begin{table}[h] 
   \caption{Success Rates in Partially Observable Environments of Increasing Size} 
   \label{tab:communication-success-rates} 
   \centering 
   \begin{tabular}{llc} 
     \toprule 
     Environment Setting & Communication Model & Success Rate \\ 
     \midrule 
     $10 \times 10$ & Baseline (No Communication) & 0.0\% \\ 
     $10 \times 10$ & Learned Direct Communication (LDC) & 30.8\% \\ 
     $10 \times 10$ & \textbf{Intention Communication} & \textbf{99.9\%} \\ 
     \midrule 
     $15 \times 15$ & Baseline (No Communication) & 0.0\% \\ 
     $15 \times 15$ & Learned Direct Communication (LDC) & 12.2\% \\ 
     $15 \times 15$ & \textbf{Intention Communication} & \textbf{96.5\%} \\ 
     \bottomrule 
   \end{tabular} 
 \end{table} 

 The baseline model, without communication, fails to learn a successful policy. This is expected and highlights the necessity of communication for this task. LDC achieves a 30.8\% success rate in the $10 \times 10$ grid, showing an emergent protocol can provide some benefit. However, its performance collapses to 12.2\% in the larger $15 \times 15$ grid. The unstructured, low-bandwidth message is insufficient for the increased complexity. 

 In contrast, Intention Communication achieves near-perfect success rates in both environments, maintaining 96.5\% performance in the $15 \times 15$ setting. This result demonstrates the scalability of an engineered, world model-based protocol. 

 \subsection{Discussion: Why Do Engineered World Models Scale Better?} 
 The performance gap is attributable to the quality and structure of the learning signal. 

 \paragraph{Decoupled Learning Problem:} Intention Communication decouples the problem. The ITGM learns a simple forward-dynamics model, while the policy learns to act given both agents' plans. This provides a richer, more stable learning signal than LDC. 

 \paragraph{Credit Assignment:} In LDC, the agent must associate a single bit with a complex action sequence from a delayed reward, making credit assignment immense. In Intention Communication, the message is tied to the imagined trajectory. Gradients can more easily shape the planning (ITGM) and summarization (MGN) modules to produce better messages. 

 \paragraph{Richness of Information:} A single bit in LDC can only resolve binary ambiguity. The MGN's vector message, grounded in a trajectory, encodes richer information like direction and destination. This allows for nuanced coordination, which is critical as complexity grows. 

 \subsection{Implications for Sample Efficiency and Embodied AI} 
 Notably, these results were achieved under significant computational constraints (Google Colab), limiting training time and batch sizes. Intention Communication's success here highlights its sample efficiency. For embodied AI, where physical interaction is costly, sample efficiency is a fundamental requirement. Our work strongly suggests that giving agents predictive world models and structuring communication around them is an effective strategy for creating sample-efficient and scalable multi-agent systems. It shows that for complex coordination, engineered biases are more effective than purely emergent protocols. 

 \section{Conclusion and Future Work} 

 We compared emergent and engineered communication for cooperative multi-agent decision-making, using the lens of embodied world models. We introduced LDC, where a protocol emerges from scratch, and Intention Communication, where agents use a learned world model (ITGM) to plan and communicate future trajectories. 

 Our results are clear. While LDC facilitates basic coordination in simple settings, it fails to scale to larger, more complex, partially observable ones. In contrast, Intention Communication, with its engineered inductive bias for prediction, shows exceptional performance, maintaining near-perfect success rates as complexity increases. This highlights the impact of world models on performance, scalability, and sample efficiency, critical factors for deploying embodied agents. 

 The central takeaway is that for robust, scalable coordination, an emergent protocol from a single, unstructured reward signal is insufficient. Instead, giving agents explicit predictive models and structuring their communication around the model outputs provides a more reliable foundation for intelligent interaction. 

 This research opens several avenues for future work: 
 \paragraph{Stochastic and Richer World Models:} Our ITGM is simple and deterministic. Future work should explore more sophisticated, stochastic world models (e.g., VAEs or Diffusion Models) that can represent uncertainty. This would enable agents to communicate plans and confidence, leading to more robust decisions under uncertainty. 
 \paragraph{From Grids to High-Fidelity Physics:} The principles from our grid-world are a proof of concept. The next step is applying these strategies to more complex embodied domains, such as robotic manipulation in simulators like Isaac Gym or on real-world robotic teams, and investigating sim-to-real transfer. 
 \paragraph{Adaptive Communication Protocols:} While our protocol is structured, the message content is learned. Future research could explore hybrid approaches where agents dynamically adjust the detail in their communicated plans. 
 \paragraph{Scaling to Larger Teams:} We focused on two agents. Scaling these models to larger, heterogeneous teams presents a complex challenge. 

 Ultimately, our findings advocate for integrating learned, predictive world models as a cornerstone of communication and decision-making for embodied AI, moving beyond a purely end-to-end philosophy. 

\begin{ack}
We thank Denon Chong Cheng Zong and Jeff Lee for their contributions to background research on MARL communication. We thank Dr. Matthew Berland for their mentorship and feedback.
\end{ack}

\section*{Contributions}
\textbf{Brennen Hill}: Project concept, project leadership, environment development, baseline model development, Learned Direct Communication development, Intention Communication development, scaling, optimization, MARL communication research, agent training.

\textbf{Mant Koh En Wei}: Intention Communication development, baseline model development, optimization, scaling, agent training.

\textbf{Thangavel Jishnuanandh}: Baseline model development, MARL communication research, optimization, scaling, agent training.

 \bibliography{main} 

\begin{thebibliography}{17}
\providecommand{\natexlab}[1]{#1}
\providecommand{\url}[1]{\texttt{#1}}
\expandafter\ifx\csname urlstyle\endcsname\relax
  \providecommand{\doi}[1]{doi: #1}\else
  \providecommand{\doi}{doi: \begingroup \urlstyle{rm}\Url}\fi

\bibitem[Anglen(2024)]{anglenMarlImpact}
Jesse Anglen.
\newblock The impact of multi-agent reinforcement learning (marl).
\newblock \emph{Rapid Innovation}, 2024.

\bibitem[Foerster et~al.(2016)Foerster, Assael, de Freitas, and Whiteson]{foerster2016learning}
Jakob N. Foerster, Yannis M. Assael, Nando de Freitas, and Shimon Whiteson.
\newblock Learning to communicate with deep multi-agent reinforcement learning.
\newblock \emph{Advances in Neural Information Processing Systems}, 29:\penalty0 2137--2145, 2016.
\newblock \doi{10.5555/3157096.3157290}.
\newblock URL \url{https://papers.nips.cc/paper/6570-learning-to-communicate-with-deep-multi-agent-reinforcement-learning.pdf}.

\bibitem[Hafner et~al.(2020)Hafner, Lillicrap, Ba, and Norouzi]{hafner2019dream}
Danijar Hafner, Timothy Lillicrap, Jimmy Ba, and Mohammad Norouzi.
\newblock {Dream to Control: Learning Behaviors by Latent Imagination}.
\newblock In \emph{International Conference on Learning Representations}, 2020.

\bibitem[Hafner et~al.(2021)Hafner, Lillicrap, Norouzi, and Ba]{hafner2020mastering}
Danijar Hafner, Timothy Lillicrap, Mohammad Norouzi, and Jimmy Ba.
\newblock Mastering atari with discrete world models.
\newblock In \emph{International Conference on Learning Representations (ICLR)}, 2021.

\bibitem[Hafner et~al.(2023)Hafner, Pasukonis, Ba, and Lillicrap]{hafner2023mastering}
Danijar Hafner, Jurgis Pasukonis, Jimmy Ba, and Timothy Lillicrap.
\newblock Mastering diverse domains through world models.
\newblock \emph{Transactions on Machine Learning Research (TMLR)}, 2023.
\newblock ISSN 2835-8856.
\newblock URL \url{https://openreview.net/forum?id=9x6o2s6KEp}.

\bibitem[Kefan et~al.(2024)Kefan, Siyan, Jiechuan, Chuang, Xiangjun, and Zongqing]{kefanMarlstationary}
Su~Kefan, Zhou Siyan, Jiang Jiechuan, Gan Chuang, Wang Xiangjun, and Lu~Zongqing.
\newblock Multi-agent alternate q-learning.
\newblock \emph{IFAAMAS}, 2024.

\bibitem[Kingma and Ba(2014)]{kingma2014adam}
Diederik~P. Kingma and Jimmy Ba.
\newblock Adam: A method for stochastic optimization, 2014.

\bibitem[Konda and Tsitsiklis(1999)]{NIPS1999_6449f44a}
Vijay Konda and John Tsitsiklis.
\newblock Actor-critic algorithms.
\newblock In S.~Solla, T.~Leen, and K.~M\"{u}ller, editors, \emph{Advances in Neural Information Processing Systems}, volume~12. MIT Press, 1999.
\newblock URL \url{https://proceedings.neurips.cc/paper_files/paper/1999/file/6449f44a102fde848669bdd9eb6b76fa-Paper.pdf}.

\bibitem[Liu et~al.(2021)Liu, Wan, Sui, Sun, and Lan]{liu2021islff}
Zeyang Liu, Lipeng Wan, Xue Sui, Kewu Sun, and Xuguang Lan.
\newblock Multi-agent intention sharing via leader-follower forest.
\newblock \emph{arXiv preprint arXiv:2112.01078}, 2021.
\newblock \doi{10.48550/arXiv.2112.01078}.
\newblock URL \url{https://arxiv.org/abs/2112.01078}.

\bibitem[Ming et~al.(2024)Ming, Kaiyan, Yiming, Renzhi, Yali, Furui, Mingliang, and Leong]{yangMarlCommsEffective}
Yang Ming, Zhao Kaiyan, Wang Yiming, Dong Renzhi, Du~Yali, Liu Furui, Zhou Mingliang, and Hou Leong.
\newblock Team-wise effective communication in multi-agent reinforcement learning.
\newblock \emph{Springer US}, 2024.

\bibitem[Mordatch and Abbeel(2018)]{mordatch2018emergence}
Igor Mordatch and Pieter Abbeel.
\newblock {Emergence of Grounded Compositional Language in Multi-Agent Populations}.
\newblock In \emph{Proceedings of the AAAI Conference on Artificial Intelligence}, volume~32, pages 871--879, 2018.

\bibitem[Raileanu et~al.(2018)Raileanu, Denton, Szlam, and Fergus]{raileanu2018modeling}
Roberta Raileanu, Emily Denton, Arthur Szlam, and Rob Fergus.
\newblock Modeling others using oneself in multi-agent reinforcement learning.
\newblock In Jennifer Dy and Andreas Krause, editors, \emph{Proceedings of the 35th International Conference on Machine Learning}, volume~80 of \emph{Proceedings of Machine Learning Research}, pages 4275--4284. PMLR, 10--15 Jul 2018.

\bibitem[Sukhbaatar et~al.(2016)Sukhbaatar, Szlam, and Fergus]{sukhbaatar2016learning}
Sainbayar Sukhbaatar, Arthur Szlam, and Rob Fergus.
\newblock Learning multiagent communication with backpropagation.
\newblock \emph{Advances in Neural Information Processing Systems}, 2016.
\newblock URL \url{https://papers.nips.cc/paper/6398-learning-multiagent-communication-with-backpropagation.pdf}.

\bibitem[Sutton and Barto(2018)]{sutton2018reinforcement}
Richard~S. Sutton and Andrew~G. Barto.
\newblock \emph{Reinforcement Learning: An Introduction}.
\newblock MIT Press, 2nd edition, 2018.
\newblock URL \url{http://incompleteideas.net/book/the-book-2nd.html}.

\bibitem[Terry et~al.(2021)Terry, Black, Grammel, Jayakumar, Hari, Sullivan, Santos, Dieffendahl, Horsch, Perez-Vicente, et~al.]{terry2021pettingzoo}
J~Terry, Benjamin Black, Nathaniel Grammel, Mario Jayakumar, Ananth Hari, Ryan Sullivan, Luis~S Santos, Clemens Dieffendahl, Caroline Horsch, Rodrigo Perez-Vicente, et~al.
\newblock Pettingzoo: Gym for multi-agent reinforcement learning.
\newblock \emph{Advances in Neural Information Processing Systems}, 34:\penalty0 15032--15043, 2021.

\bibitem[Vaswani et~al.(2017)Vaswani, Shazeer, Parmar, Uszkoreit, Jones, Gomez, Kaiser, and Polosukhin]{vaswani2017attention}
Ashish Vaswani, Noam Shazeer, Niki Parmar, Jakob Uszkoreit, Llion Jones, Aidan~N. Gomez, {\L}ukasz Kaiser, and Illia Polosukhin.
\newblock {Attention Is All You Need}.
\newblock In \emph{Advances in Neural Information Processing Systems}, volume~30, pages 5998--6008, 2017.

\bibitem[Vincent(2024)]{vincentMarlRobotics}
Caroline~R Vincent.
\newblock Multi-agent reinforcement learning for autonomous robotics.
\newblock \emph{DSpace@MIT}, 2024.

\end{thebibliography}
 \bibliographystyle{plainnat} 

 \appendix 

 \section{Technical Appendices and Supplementary Material} 

 \subsection{Exploration of LDC Fragility and Design Choices} 
 We conducted supplementary experiments to better understand LDC's limitations and sensitivities. These findings underscore the difficulty of fostering stable emergent communication without strong priors, especially under computational constraints. 

 \paragraph{Varying the Message Space Capacity:} 
 We hypothesized that the limited bandwidth of a single binary message might be the bottleneck for LDC. We expanded the message capacity by increasing the range of a single token and by allowing multiple tokens per timestep. Surprisingly, as shown in Table \ref{tab:convergence}, only configurations with a very low-dimensional message space (a single token with a range of 0-1 or 0-2) reliably converged to a success rate >1\%. 

 We theorize that a larger message space dramatically increases the dimensionality of the joint action-observation space. This adds noise to the learning signal and exacerbates the credit assignment problem, making convergence on a coherent protocol from a sparse reward nearly impossible. 

 \begin{table}[h]
  \caption{Convergence Results for LDC with Varying Message Capacity}
  \label{tab:convergence}
  \centering
  \begin{tabular}{ccc}
    \toprule
    Message Range & Message Count Per Step & Converged to $>$1\% Success Rate \\
    \midrule
    0-1  & 1  & \textbf{Yes} \\
    0-1  & 2  & No  \\
    0-1  & 4  & No  \\
    0-1  & 10  & No  \\
    \midrule
    0-2  & 1  & No \\
    0-2  & 2  & No  \\
    0-2  & 4  & No  \\
    0-1  & 10  & No  \\
    \midrule
    0-4  & 1  & No  \\
    0-4  & 2  & No  \\
    0-4  & 4  & No  \\
    \midrule
    0-9  & 1  & No  \\
    0-9  & 2  & No  \\
    0-9  & 4  & No  \\
    \midrule
    0-99  & 1  & No  \\
    0-99  & 2  & No  \\
    0-99  & 4  & No  \\
    \bottomrule
  \end{tabular}
\end{table}

 \paragraph{Environment Observation Variations:} 
 We experimented with giving agents absolute, rather than relative, grid coordinates, hoping for a more stable frame of reference. However, LDC policies consistently failed to converge. We suspect this is because deriving directional information from absolute coordinates requires extra computation, increasing the policy function's complexity. The increased variance in observations likely hindered learning a stable policy. 

 \paragraph{Explicitly Rewarding "Meaningful" Messages:} 
 To guide protocol learning, we tried adding an intrinsic reward for "meaningful" messages, based on a message's influence on the recipient's value function estimate. However, this reward shaping was difficult to tune. High coefficients destabilized learning the primary task, while low coefficients had no effect. This highlights the challenge of designing auxiliary rewards that align with the main task. 

 \paragraph{Forcing the Need for Communication:} 
 To create a scenario where communication was indispensable, we designed an environment where each agent could only see its teammate's goal. Success is impossible without information exchange. Even here, LDC failed to converge, further showing its limitations in discovering strategies from scratch. 

 \paragraph{Network Architecture Variations:} 
 We explored various policy/value network architectures. Deviations from our standard architecture (two 64-node hidden layers), whether larger or smaller, generally resulted in longer training times or failure to converge. This suggests emergent communication methods can be highly sensitive to hyperparameters, adding to their practical difficulty. 

 \subsection{Conclusions from Additional Experiments} 
 These experiments show emergent communication is a powerful but fragile phenomenon. Without structured environments, rewards, and biases, effective protocols are exceedingly difficult to emerge, especially under typical computational and sample constraints. The consistent success of Intention Communication highlights the value of embedding structured, world model-based priors into agents to achieve robust and sample-efficient learning. 

\subsection{Key Training and Architectural Hyperparameters}
We share the hyperparameters used in this experiment for greater reproducability in table \ref{tab:hyperparams}.

\begin{table}[h]
\caption{Key Training and Architectural Hyperparameters}
\label{tab:hyperparams}
\centering
\begin{tabular}{lc}
\toprule
Parameter & Value \\
\midrule
\multicolumn{2}{l}{\textit{Training Hyperparameters}} \\
Initial LR ($\text{lr}_0$) & $1 \times 10^{-3}$ \\
Final LR ($\text{lr}_f$) & $1 \times 10^{-4}$ \\
Discount Factor ($\gamma$) & 0.99 \\
Optimizer & Adam \citep{kingma2014adam}\\
Critic Loss Coefficient ($\alpha_c$) & 0.5 \\
Max Episode Length ($T_{max}$) & 200 \\
\midrule
\multicolumn{2}{l}{\textit{Architectural Hyperparameters (all)}} \\
Frame Stack ($F$) & 4 \\
Hidden Dimension ($D_{hidden}$) & 64 \\
Hidden Layers ($N_{hidden}$) & 2 \\
Message Loss Coefficient ($\beta$) & 0.1 \\
\midrule
\multicolumn{2}{l}{\textit{Architectural Hyperparameters (Intention Communication)}} \\
ITGM Rollout Horizon ($H$) & 3 \\
Attention Heads ($N_{heads}$) & 1 \\
Message Components ($N_{comp}$) & 8 \\
Symbol Vocabulary Size ($N_{sym}$) & 8 \\
\bottomrule
\end{tabular}
\end{table}

\end{document}